%Paper: hep-th/9506178
%From: chemtob@amoco.saclay.cea.fr (Marc Chemtob)
%Date: Tue, 27 Jun 1995 11:28:42 +0200
%Date (revised): Tue, 18 Jul 1995 18:53:33 +0200

%\magnification\magstephalf
\magnification\magstep 1
\parskip = 0.2 true cm
\baselineskip = 0.5 true cm
%\baselineskip = 0.847 true cm
\overfullrule =0 pt
\hfuzz = 1 pt

\def \GUT { grand unified theories }
\def \wrt { with respect to }
\def \cft { conformal field theory }

\def \bet { beta function }
\def \mow { modular weights }
\def\loe { low energy }

\def \susy { supersymmetry }
\def \susyq { supersymmetric }
\def \sig { $\sigma -$model }
\def \ws { world sheet }
\def \tres { threshold corrections }
\def\L {\Lambda }
\def\l {\lambda }
\def \t {\theta }
\def \vt {\vartheta }
\def\a {\alpha }
\def\dh {\partial }
\def \d {\delta }
\def \D {\Delta }
\def \bq {\bar q }

\def \g {\gamma }
\def \G {\Gamma }
\def \O {\Omega }

\def \b {\beta }

\def \s {\sigma }
\def \e {\epsilon }
\def \cc { coupling constant }
\def \ccs {coupling constants }

\def \ren { renormalization }

\def \ud { {1 \over 2} }

\def \sm { standard model }
\def \q {\eqno  }
\def \mssm { minimal supersymmetric standard model }
\def \he  { high energy }
\def \st { string theory }
\def \ft { field theory }

\rightline {hep-th/9506178}
\rightline  {T95/086}
\vskip 1.cm
\centerline {\bf  THRESHOLD CORRECTIONS IN ORBIFOLD MODELS }
\vskip 0.5cm
\centerline {\bf  AND SUPERSTRING UNIFICATION OF GAUGE INTERACTIONS }

\vskip 2cm
\centerline {M. CHEMTOB \footnote \dag
{\it e-mail: chemtob at amoco.saclay.cea.fr }}
\vskip 1cm
\centerline {\it Service de Physique Th\'eorique}
\centerline {\it CE-Saclay}
\centerline {\it F-91191 Gif-sur-Yvette Cedex France
\footnote \ddag {\it Laboratoire de la Direction des Sciences
de la Mati\`ere du Commissariat \`a l'Energie Atomique }}

\vskip 1cm
%{\bf ABSTRACT }

The string one loop \ren  of the gauge
\ccs is  examined   in abelian
orbifold  models. The
contributions to  string
threshold corrections independent of the compactification
moduli fields are   evaluated  numerically for several
representative examples of  orbifold models.
We consider  cases with
standard and  non-standard embeddings as well as cases with
discrete Wilson lines
background fields
which match reasonably well
with \loe
phenomenology.
 The   gap separating
the observed  \GUT  scale $M_{GUT}\simeq  2\times 10^{16} GeV $
from the string unification scale $M_X \simeq  5\times 10^{17} GeV$  is
discussed on the basis of standard-like  orbifold
models.
We examine  one loop gauge \ccs unification in a description
incorporating the combined effects of  moduli dependent and independent
threshold corrections, an adjustable affine level for the hypercharge group
factor and a large
mass threshold  associated with an anomalous $U(1)$
mechanism.
\vskip 2cm
{\bf PACS numbers:} 11.10.Hi, 11.25.Mj, 12.10.Kt
\vfill \eject

{\bf 1. INTRODUCTION  }

\vskip 1cm
In superstring as in grand  unification, the high energy extrapolation
of  the \sm  renormalized gauge \ccs
is described by a
one loop scale  evolution of  familiar form:
$${(4 \pi )^2\over g_a^2(\mu ) } =
{(4 \pi )^2k_a\over g_{X}^2} +2b_a \log {\mu
\over M_{X}}+\tilde \Delta_a (M_i,\overline  M_i). \q (1)$$
(The index  $a=3,2,1 $ labels the $SU(3)\times SU(2)\times U(1)$
group factors $G_a$ and
$b_a$ are  the  \bet  slope parameters associated with the  \loe modes,
 $\b_a (g)=-b_a g_a^3/(4\pi )^2 +\cdots $)
The superstring case is, however,
distinguished by three important features [1]:

(i) Tree level relations [2] involving the gauge and
gravitational interactions,
$$g_{X}^2= k_a g_a^2= {4 \kappa ^2\over \a'}
={32\pi \over\a' M_P^2}
 . \q (2)$$
  In addition to the
string theory expansion parameter $g_X$  (or  4-d  dilaton
VEV   $<S>=1/g_X^2$) which is specified  by the ratio of the
string  mass scale  $M_S={2\over \sqrt {\a'}}$ to  the
phenomenological Planck mass $M_P ={\sqrt {8\pi } \over \kappa }
=1.22 \times 10^{19} GeV$,
as exhibited  in eq.(2),
three  extra  free  (positive rational numbers)
parameters $k_a$ are introduced in eq.(2), corresponding to the
levels of the affine Lie algebras for the  gauge group
factors $G_a$ in the
underlying string theory.

(ii) An   improved unification scale $ M_X$ defined in  eq.(1)
as the  matching  scale between the field
and string  theories
renormalized \ccs at which   these   obey
most closely the tree level relations, eq.(2).
For the  \ft    \ccs in the $ \overline {DR} $  regularization  scheme,
$$M_{X}= {e^{(1-\g )/2} \over 4\pi \root 4 \of {27}}  g_{X}
M_P
= {e^{(1-\g )/2} \over  \root  \of {2 \pi } \root 4 \of{27}}  M_{S}
\simeq g_{X} 5.27 \times 10^{17}  GeV.   \q (3) $$
The \ft (ft) convention    in  use here
is related to the \st  (st) one  as,
$g_a^{ft}=\sqrt 2 g_a^{st} $, corresponding to the normalization
of the  Lie algebra generators,
 $ Tr_R(Q_a^2)=
 \ud c(R)$,  where $c(R)=l(R)$ is the   Dynkin index of representation $R $.

 (iii) Threshold corrections
accounting  for the
contributions of the infinite set of massive string states
at the string ($M_S$) and   compactification ($M_C$) scales, integrated out
by matching the field and string theories scattering amplitudes.
These corrections are represented in eq.(1)
by  the functions $\tilde \D_a (M_i, \overline M_i)$  depending  upon
the structure of the string  mass spectrum and the other
characteristic  parameters of the compactified space manifold, such as
the  vacuum expectation values (VEVs) of the compactification
moduli fields, $ M_i =T_i, U_i$ [3].
Specifically, $M_X$
is defined as the choice of scale  which
minimizes  the \tres contributions.
Of course, the perturbative character of formula (1) implies that
the size  of $\tilde  \D_a$ should be  comparable to
that of two-loop effects, so  that  $\tilde \D_a = O (1)$.

For a quantitative test
of superstring unification based on eq.(1)
and  for a proper identification of the fundamental
parameters, $M_X $ and $g_X$,
it is essential to understand the structure and
size  of threshold corrections.
Thus, an additive decomposition  such as
 $ \tilde \D_a =k_a Y -b_a \D
$,   may  be exploited to introduce  effective unification scale
and coupling constant,
$$M_{X}\to M'_{X}
=M_{X} e^{\D /2},\quad
 g_{X}
\to g'_{X}={g_{X}\over
(1+{Y g_{X}^2\over (4\pi )^2})^{\ud } }
,\q (4)$$
so defined as to  incorporate the contributions
from the above two components,  $Y $ and $\D  $.

 The toroidal compactification    orbifold models
 prove very helpful in obtaining an information on $\tilde \D_a$.
  The contributions from  compactification modes
admit here   a   natural additive  decomposition into
a moduli dependent  component
  arising from  the
  chiral mass F-terms and a moduli independent component
  arising from the  vector mass D-terms [4]. As is well known,  the
  moduli dependent  contributions play an essential r\^ole in the
  cancellation of sigma model anomalies  affecting  the
  target space duality  symmetry [5].
  These  can be represented  by general formulas involving
  the automorphic functions of the compactication manifold  accompanied by
  model dependent coefficients.
  On the other hand,  the
  moduli independent contributions carry  only an implicit dependence on
  the compactification manifold, such as the orbifold gauge embedding
  or the discrete  Wilson lines.
  In spite of several attempts in the literature
  to estimate  the size of both components of \tres [1,6-9],
  one is still lacking  a clear physical
  understanding of  their magnitude.
Our main goal in this paper   is to present  results  for
the moduli independent  \tres
through an extensive  numerical study based on
a sample of orbifold models.
A recent work by Dienes and Faraggi [34], which appeared while this paper
was being completed, pursues a similar goal to ours based on the
fermionic models.

The main
physical motivation for this paper  is, however,
the wide  gap that separates  the  improved
string unification scale  $M_X = 0.216 M_S\simeq 5 \times
10^{17} GeV $, assuming $g_X =O(1)$,
from the  observed grand  unification  scale, $M_{GUT}\simeq
2\times  10^{16} GeV$, as
determined by extrapolating the gauge \ccs up from their
 experimentally determined  values at  the   $Z-$boson  mass  [10].
 The  implications of this order of magnitude  discrepancy
 in scales have  been emphasized on
 several occasions [11]. The conflict for  superstring unification
 can be resolved in two different ways:
One can postulate [12]  large
string   threshold corrections such that after
becoming equal and joining  together at the observed   scale  $M_{GUT}$
the  gauge \ccs  follow diverging flows
up to   $M_X$.
A matching of  the one loop  extrapolated values  of $g_a(M_X)$
with their  predicted  values,  as obtained  by adjusting the
moduli dependent  threshold corrections,
can be successfully  achieved in terms of
wide classes of solutions
for the  \mow of massless modes consistent with  the anomaly
cancellation constraints [5,12].
Alternatively, one can postulate  [13] an
affine level parameter  for  the
weak hypercharge group $U(1)_Y$ somewhat  lower than
the standard  grand unification group value, $k_1= {5\over 3} $.  With
such an enhanced starting
value  for
$ (k_1 \a^2_1 (m_Z))^{-1} $ one achieves a  delayed joining
of the gauge \ccs  flows which   can easily
raise up the unification scale
by one order of magnitude.
While either of these possibilities
is well motivated  by itself and
appears sufficient to
rescue a superstring grand desert scenario, there
remains certain  unsatisfactory points.  Thus,
the  VEVs of moduli fields requested in
the first possibility, $<T>= 10-30 $,
appear to be   somewhat too  large,
whereas  no  known realistic  orbifold example [14,34]  exists  for which
the hypercharge group level parameter comes as low as the
 value $k_1 \simeq 1.4$  favored  in the second possibility.

A generic feature  of  standard-like orbifold models
is the occurrence of a rich  spectrum of charged massless
modes    appearing
 on side of the requested  (quarks  and leptons)
 chiral families  in vector representations
 of  the color and weak  groups. In fact, the matter  representations
 of the observable sectors group factors
 are generally sizeable enough so that the
 corresponding  \bet parameters  $\b_a$
 arise with  either    small negative  values   or   large
 positive  values.
 This suggests  that a first stage of  slow
or non asymptotically free
scale  evolution  may well
take place from $M_X $ down to some scale
where   the extra  modes
 pair up by acquiring mass  and decouple.
 As is well-known [15], in order for the 4-d \loe effective  theory  to be
 weakly coupled,  so as not to invalidate the use of eq.(1) ($g_X
 \approx g_{d} M_C^3 < 1 , \quad g_d= $
 10-dimensional  gauge coupling constant),
 and in order
 to avoid dealing with a strongly coupled
 10-d theory ($ g_d M_S^3 <1 $) one must require that
 the compactification and unification scales  retain a
  magnitude comparable  to  the string scale,  $ M_X \simeq M_C \simeq M_S$.
  (The  second restriction can be relaxed by allowing, for instance,
  for an anisotropic compactification manifold (large radius in one
  out of the  six  compactified dimensions)  in which
  a weakly coupled effective theory,
  $g_X <O(1)$, could remain compatible
  with a strongly coupled string theory
  (large $g_d$) [16].)
 Assuming  the above near equality of scales,
 a natural identification  for the decoupling scale of the  extra matter
 is the  mass  scale, denoted $M_A$, which is  induced
 by a non-vanishing Fayet-Iliopoulos D-term contribution to some
 apparently anomalous $U(1)$ group factor
 occurring  on compactification [17].
 This suggestion is not new, of course,
 and appears in several places in the specialized literature. The idea is
to cancel  the non vanishing one loop string
contributions  to the D-term
scalar potential of an apparently anomalous $U(1) $ factor by judiciously
lifting the VEVs of certain
scalar fields while restoring a stable
\susyq  vacuum.
We shall carry out an analysis of the  one loop gauge \ccs unification
which combines  together the above ideas of adjustable
moduli VEVs and $k_1$ level parameters together
with that of  an adjustable  intermediate
scale $ M_A$, while  describing the scale evolution  in the interval
from $M_X $ to $ M_A$  on the basis of
orbifold models predictions.

The paper contains 5 sections.
In Section 2, we discuss in wide outline
the basic formalism involved in the
one loop string renormalization of the gauge coupling
constants as applied to
orbifold models. None of the results discussed in this section is new, our
main intent being to provide a concrete, encapsulated presentation of
the relevant formalism.
In Section 3, we  present numerical results for the moduli independent \tres
for a sample of representative orbifold models.
In Section 4, we  examine  the viability of superstring unification
in an extended picture including \tres and an intermediate scale
associated with an anomalous $U(1)$ symmetry.
In Section 5, we summarize the main conclusions.

\vfill\break
{\bf 2. ONE LOOP STRING RENORMALIZATION  }
\vskip 1cm

{\bf 2.1 Threshold corrections to gauge coupling constants}
\vskip 0.5cm

We consider the class of
\loe  \susyq  theories
descending  from
4-d heterotic string theories
with a  nonsemi-simple gauge group
$\prod_a G_a $.
The  genus  zero  (unity) \ws   (with Wick-rotated
Euclidean metric)  of the \cft is a  sphere  (torus)
parametrized by  planar  coordinates:
$ \bar z =  e^{-2\pi i\bar \zeta } ,
z=  e^{2\pi i \zeta } $, with  corresponding
cylindrical coordinates given for the sphere by:
$\bar \zeta = \s  -i t , \quad
\zeta = \s  +i t   , \quad  \s \in [0,1], \quad t\in [-\infty , \infty ]  $
and for the torus by:
$ \zeta = \s +\tau t ,
\quad \bar \zeta = \s +\bar \tau t , \quad
\s , t\in [0,1], \tau =\tau_1+i\tau_2   $. The
right-moving  RNS (Ramond-Neveu-Schwarz) superstring is
built  with   20  spacetime and spin
fields
$X^\mu (\bar z) , \psi^\mu
(\bar z), \quad
[\mu =0,\cdots , 9] $, associated with
$D=4 $   external dimensions of the
flat spacetime $[\mu =0, \cdots , 3]$     and $ d-D=10-D=6 $ internal
dimensions $[ \mu =4,\cdots 9]$ of the compactification space manifold,
 represented in a complex basis as,
$X_R^i, X_R^{\bar i} , \quad  \psi^i =e^{i\phi_i} ,
 \psi^{\bar i}=e^{-i\phi_i}
, [i=1,2,3]$
where the complex  scalar fields  $\phi_i (z)$ are   coordinates
of the $SO(6) $ group Cartan torus. This is
tensored  by
a left-moving  bosonic  string
built  with 26  fields  $X^\mu (z) , [\mu = 0,..., 25]$,
comprising  $ D $ external  space  coordinates
and  $ 26-D $  internal space coordinates  which are distributed
into 6  compactified space   coordinates
$X_L^i, X_L^{\bar i}  $
and  16 gauge  coordinates  of the  $E_8\times E_8'$ Cartan  torus
$F^I, F^{'I} [I=1,...,8]$
generating the currents $J_a (z)$ of the
affine Lie algebras  $G_a$
 of  levels  $k_a$. At certain places, we refer to these
coordinates globally as $F^I, [I=1,\cdots , 16] $
and   also by using their  fermionic
representation in terms of  complex  2-d Weyl spinors,
$(\l^\a , \l^{\bar \a } )= e^{\pm iF^I},
[I=1,\cdots , 16; \quad \a =1,\cdots 8
]$. Of course, the above covariantly quantized string theory  must
be supplemented with the
conformal ghost fields $c^z(z,\bar z), b_{zz}(z,\bar z)$  and the
superconformal  ghost fields, $ \g (\bar z) ,  \b_z (\bar z) $ [18].

The one loop  string \tres are described by a   general formula
obtained by Kaplunovsky [1],
$$\tilde \Delta_a\equiv k_aY_0 +\D_a,\qquad \D_a  =
-\int_F {d^2\tau\over  \tau_2 }
\bigg (k_a B_a (q,\bq ) - b_a  \bigg ), \q (5)$$
where one  has decomposed the total contribution, denoted
$\tilde \D_a$, into  a universal  contribution,
$k_a Y_0$,  independent of the
gauge group  factor (except for the coefficient $k_a$),
arising  from gravitational interactions
 and oscillator excitations  modes, and a  contribution solely due to
the massive compactification modes, denoted $\D_a$. The latter
component  is expressed as a  deformed partition function
integrated over
the inequivalent complex structures of  the
 genus 1  world sheet, with an integrand
$$\eqalign {B_a (q ,\bq ) =&-
\ud \sum_{ even\quad  \bar \a , \bar \b }
\bigg [(-1)^{2\bar \a + 2\bar \b }
{1 \over \eta(\tau )^2 \eta (\bar \tau )^2}
2\bq {d \over d\bq } \bigg (
{r \vt \bigg [{\bar \a\atop \bar \b }\bigg ]
(\bar \tau ) \over \bar \eta (\bar \tau ) } \bigg ) \bigg ] \cr
& \times 2Trace \bigg ( (-1)^{2\bar \b F } Q_a^2  q^{L_0-{22\over 24} }
\bar q^{\bar L_0 - {9\over 24}}\bigg ) , }\q (6)$$
where the  first factor represents the partition function
of the external theory inserted
 with the  operator   $(-{1 \over 12} +\chi^2 )$, where
$ \chi $ denotes the  4-d helicity or chirality  vertex operator and we
have introduced the familiar Dedekind function,
$\eta (\tau )=q^{1\over 24} \prod_n (1-q^n) $, and
the Jacobi theta-functions (cf. eq.(10) below).

The second factor in eq.(6) (with
$F=$ fermion number operator, $L_0, \bar L_0$ =
conformal dimensions operators)  corresponds to
the internal  theory partition function
inserted with   the square $Q_a^2$  of any
one of the  gauge group  generators for
subgroup $G_a$.
  The integral over
 the \ws   torus complex
 modular parameter,
  $\tau =\tau_1+i\tau_2 $,  with  $q =e^{2\pi i \tau } ,
  \bar q =e^{-2\pi i\bar  \tau } ,  $
extends  over the
modular  group $SL(2,Z)$ fundamental   domain, $F=[\vert
\tau_1 \vert \le \ud , \vert \tau
\vert \ge 1 ]$.
Infrared  convergence of the integral, eq.(5),
is ensured by the subtraction of
  $b_a=\lim_{\tau_2\to \infty }
 k_a B_a$,   where $b_a ={1\over 6} \sum_\a [-c_S(R_\a )-2c_F(R_\a )+
 11c_V(R_\a )] $
 ($S=$ complex scalar, $F=$  Weyl or Majorana fermion, $V=$ vector)
 represent the summed  contributions to the \bet slope parameters
 from the  massless string modes $\a $.

 The summation  in eq.(6)
 over  the subset of  even  spin structures of the
 right-moving sector, $ (\bar \a , \bar \b ) = [(0,0), (0,\ud ),
 (\ud ,0) ]$ where    $ \bar \a , \bar \b =   0=NS  (A)  $
 (Neveu-Schwarz,    Antiperiodic)   or $  \ud = R (P)  $ (Ramond, Periodic)
 is performed by insertion of the familiar GSO
 (Gliozzi-Scherk-Olive) projection
 phase factors leading  to the \susyq string  [18].

\vskip 0.5cm
{\bf 2.2  Specialization to orbifolds}
\vskip 0.5cm

 To express the  second  internal  space
 factor in eq.(6) for orbifolds, we recall
 first that the  projection (modding) \wrt the  orbifold point symmetry
 is achieved by summing  over the (space and time)   twisted subsectors
 $(g,h)$  by using [19,20],
$$ Trace(\cdots )  = {1\over \vert G\vert }
\sum_g \sum_{h; [g, h]=0} \chi (g,h)
Trace_g (h \cdots ),$$
 where $\vert G\vert $ is the orbifold point group order and $\chi (g, h)$
 are degeneracy factors.
 For toroidal  compactication,  all  fields are free so that
 the torus partition function
is  obtained   by associating to
a  complex coordinate field  $X(\s , t)$ of given chirality, a factor
$ 1/(\pi \sqrt { 2 \tau_2 }) $ (flat case) or
$(1-e^{2\pi i v} )/\eta (\tau )$
  (untwisted case with time twist
  $X(\s ,t+1 ) =e^{2\pi i v}X(\s , t ) $ ) or
$\eta  (\tau )/ \vartheta  [{\ud +v \atop  \ud +v} ] $
(space twisted case
$X(\s +1 , t)
= e^{2\pi i v} X(\s , t)$) and to a fermionic Majorana-Weyl
field, obeying the twisted boundary conditions:
$$ \psi (\s  +1,t)=-e^{2\pi i \t' } \psi (\s ,t ),
 \quad \psi (\s ,t+1  )=-e^{-2\pi i \phi' } \psi (\s ,t ),$$
a factor  $ [\vartheta [{\t '\atop  \phi '} ]/\eta (\tau ) ]^\ud  $.
The  zero modes are associated a factor   $ q^{p_L^2/2 } \bar q^{p_R^2/2} $
summed over the  winding modes spanning the compactification manifold
lattice $\L_6$ with basis
vectors $e_a^i$
and over  the   Kaluza-Klein momentum modes
spanning  its dual lattice $\L^\star $ with
basis vectors $ e^{\star a}_i $ (cf. eq.(11) below)).

We  recall next   that  a torus $R^6/\L^6,  $ defined by
$ X^i \equiv X^i +2\pi n^ae_a^i $, having
 a point  symmetry group, $P=Z_N$,  of automorphisms of the lattice  $\L^6$,
defines an abelian orbifold endowed  with a space
symmetry group, $ G=P\times \L^6$. The space group action on the
string  theory fields is described in terms of
 rotations $\t^k $ and translations $u_{k,f}$
 together with their associated gauge group  shift
embedding elements  described by translations
$V^I $ and Wilson lines translations $ a_a^I,  [I=1,\cdots , 16; a=1,2,3]$.
The space  group $ G =\{ g_p \} =\{ \b_p , w_p \} $
composition laws read:  $ g_1 g_2 =(\b_1 \b_2, \b_1 w_2 +w_1),
\quad g_p^{-1}= (\b_p^{-1}, -\b_p^{-1} w_p) $.

The string  Hilbert space of states consists of the
untwisted sector  ($k=0$) and the  twisted
($k=1,\cdots , N-1 $) sectors.
The twisted sectors $g^k$  are distinguished  by the  boundary conditions:
$ (X(\s +1 ,t), \psi (\s +1 , t))= g^k
(X(\s ,t),- (-1)^{2 \bar \a } \psi (\s ,t))$. They
are organized into conjugacy classes of the space group
with representative elements, $g^k =[\t^k, u_{k,f}]$ and their
associated classes,
$ \{  g^k \simeq  g'g^kg^{'-1}
=(\t^k, u_k), g'= g^p \in Z_N \} , $ where the set of shift vectors
$u_k=( \t^p u_{k,f}+(1-\t^k)u ),  \quad [u\in \L_6  , p=0, \cdots N-1 ], $
span  lattice cosets (labelled by the index $f$)
with representative elements,   $u_{k,f}$.
The compactified space coordinates, $ X^i=X_L^i+X_R^i =
x^i+i \pi tp^i+2\pi \s w^i +\cdots $ (units $2\a '=1$),
admit the  (zero and oscillators) modes
expansion,
$$(X^i_L (z) , X_R^i (\bar z)) = {x^i\over 2 }
-{i\over 2} (p_L^i \ln z , p_R^i \ln \bar z ) +{i\over 2}
\sum_{m_i} ( {\a_{m_i}^{Li}\over m_i} z^{-m_i},
{\a_{m_i}^{Ri}\over m_i} \bar z^{-m_i}).$$
In twisted sectors, the string center of mass coordinates $x^i$
are not arbitrary real parameters  but rather must satisfy:
$g^kx =x+\hat u_{k,f}+u  , \quad  [ \hat u_{k,f}, u\in \L_6 ] $.
Therefore,  each of the $g^k$  twisted sectors  splits
into subsets which can be
classified in terms of the corresponding  set
of fixed points  of the space group,  $ f^{(k)i} , $
defined as:
$ \t^k f^{(k)}  = f^{(k)} +\hat u_{k,f}$ where
$\hat u_{k,f}^i= m_{k,f}^ae_a^i,  [ m^a =  $  integers] are
translation  vectors of the 6-d
toroidal lattice $\L_6$
determined by the condition that they return
the rotated fixed point $\t^k f$ back  to its original position,
so that $f=(1-\t^k)^{-1}\hat u_{k,f} +u $.
Specifically, the k-twisted sector fixed points  $f^{(k)}_\a $
are  distinguished
by a label $\a $ running over the number of fixed points.
The lattice vectors $\hat u_{k,f}$ identify with the  lattice coset
representatives $u_{k,f}$ introduced above
only for prime orbifolds. For simply twisted
sectors, $k=1  $ or $ k=N-1  =-1 (mod N) $, the fixed points
$f^{(k)} $ and conjugacy classes $ u_{k,f}$
are in one to one correspondence, so that $f^{(k)}$ faithfully label these
classes  and $\hat u_{k,f}= u_{k,f}$. This property holds  true for
all the twisted sectors in the  prime orbifolds, $Z_{3,7}$.
For the multiply twisted sectors,
the full set of fixed points $f^{(k)}_\a $ decomposes into disjoint subsets
$\{ f_A^{(k)} , f^{'(k)}_A , \cdots  \} $,  where the fixed points
within each subset (labelled by $A$) are related as,
$\t^{p_A} f_A^{(k)} = f_A^{'(k)} \ne
f_A^{(k)} $ for $ p_A < k$, and hence are
in one to one correspondence with the same conjugacy classes, $u_{k,f}$.
The cases involving  non trivial  subsets  $[f_A^{(k)} ]$, comprising more
than one fixed point, arise
only for the non prime  ($N=1$) orbifolds  $ Z_{4,6,8,12}$
and  for the direct product orbifolds
$Z_N \times Z_M$.

The orbifold space group elements  can now be expressed as ,
$$ {g^k} = \{\t^k, u_{k,f}= m_{k,f}^ae_a; \quad
k\tilde V^I= kV^I+m_{k,f}^aa_a^i \}, $$$$  \t^k= diag (\t_i^k)
= diag(e^{2\pi ik v_i} ) .
\bigg [\sum_i v_i = 0 \bigg  ]\q (7) $$
The orbifold group action on fields (eq.(8)) and state vectors (eq.($8'$))
reads   in obvious notations:
$$  g^k X_{L,R}^i=  \t_i^k X_{L,R}^i+ 2\pi m_{k,f}^ae_a^i,\quad
g^kF^I = F^I+ 2\pi (kV^I+m_{k,f}^aa^I_a  ),
\quad  g^k\psi_i  = \t^k_i \psi_i ,\q (8)$$
$$g^h [(\a_{-n_i}^i)^{p_i}
[( \a_{-m_j}^{\bar j})^{q_j}]_{L\choose R} \vert p_R,
r^i \equiv \a^i+kv^i >_R
\vert p_L,  P^I \equiv W^I+k\tilde V^I >_L $$$$
=e^{2\pi i kh(v\cdot r   + \tilde V    \cdot P)\mp 2\pi i h (n_i +m_j)}
[ (\a ^i)^{p_i}_{-n_i} ( \a^{\bar j})^{q_j}_{-m_j}]_{L\choose R}
\vert p_R, r^i>_R \vert p_L, P^I>_L. \q (8')$$
The above  used correspondence between Wilson lines
translation vectors
and the  non contractible loops, $u_{k,f}$, refers to  abelian orbifolds.
Non abelian orbifolds with shift gauge embeddings
can be constructed by extending the definition of Wilson lines
to class dependent  shift vectors, $ k\tilde V^I \to V^I_{k,f}$
derived  from a gauge embedding matrix  of general form [21].

The  internal space oscillator operators,
$ ( \a^i_{n_i} ,  \a^{\bar j}_{m_j})_{L \choose R} $,
where  $i,  \bar j $ are complex conjugate  bases indices,
(given by the familiar linear combinations of real basis indices,
$\mu = (1+i2)/\sqrt {2},
(1-i2)/\sqrt {2},\cdots $),
enter with  the moddings,  $ n_i\in Z\mp \t_i, \quad m_j\in Z \pm \t_j $,
where $Z$ designates the set of integers.
The translation vectors $\a^i= n_i, (n_i+\ud) ,
\quad [n_i\in Z,  \sum_i n_i \in 2Z +1  $(odd  integers)] are
elements of the $SO(6) $  group  weight lattice $\G_6$ and $  W^I =n^I ,
(n^I+\ud ),  \quad [ n^I \in Z ,
\sum_{I=1}^8 n^I\in  2Z  $ (even integers) ]
are elements of the
$E_8\times E'_8$  group weight  lattice, $ \G_{8+8}$.  The translation
vectors $v^i $  and $ V^I,  a_a^I $ \wrt these lattices
must obey:
$N v^i \in \G_6,  NV^I \in \G_{8+8}, N m^aa_a^I \in \G_{8+8} $ as well as
the level matching (modular invariance under $T^N$)  conditions
$ N[(k V^I +m^a_{k,f} a_a^I)^2
-(kv^i)^2 ] \in 2 Z$.

With  the  above rules in hand,  we can now quote  the following more explicit
formula derived from eq.(6):
$$\eqalign { B_a(q,\bq )  = &- 2{1\over \vert  G\vert }
\sum_{m,n}\chi (m,n) \e (m,n)
\ud  \sum_{ even \bar \a , \bar \b }
\bigg [(-1)^{2\bar \a +\bar 2\bar  \b }
{1 \over \eta^2(\tau ) \eta^2 (\bar \tau )}
2\bq {d \over d\bq } \bigg (
{\vt \bigg [{\bar \a\atop   \bar \b } \bigg ](\bar \tau )
\over \bar \eta (\bar \tau ) } \bigg ) \bigg ] \cr
& \times
 \prod_{i=1,3} \bigg [ {\vt \bigg [{\bar \a +mv_i \atop
\bar \b +nv_i}\bigg ] (\bar \tau )
\over \bar \eta (\bar \tau ) }\bigg ]\prod_{i=1,3}\bigg [
{\bar \eta (\bar \tau ) \over \vt \bigg [{\ud +mv_i \atop
\ud  +nv_i}\bigg ] (\bar \tau )}
 {\eta (\tau ) \over \vt \bigg [{\ud +mv_i \atop  \ud
 +nv_i}\bigg ](\tau )}\bigg ] \cr & \times
 {1 \over 4} {1 \over  \eta^{16}(\tau )} \bigg [\sum_{\a, \b;\a'\b'}
 \eta (m,n; \a,\b;\a',\b')\prod_{I=1}^8 Q_a^{I2}
 \vt \bigg [{\a +m\tilde V_I \atop   \b+n\tilde V_I}\bigg ](\tau ) \cr &
 \times
 \prod_{I=1}^8 Q_a^{I'2}\vt \bigg [ {\a' +m\tilde V'_I \atop
 \b'+n\tilde V'_I} \bigg ](\tau ) \bigg ]
 \bigg [\sum_{\L_6, \L^\star _6} \ q^{p_L^2/2 } \bar q^{p_R^2/2} \bigg ]
 , }\q (9) $$
where the
second and third  factors, recognizable by the brackets,   are
contributed by the    internal space coordinates and  spinors,
 the fourth factor  by the   gauge coordinates and the last (fifth)  factor
by the compactified space  zero modes.
The numerical  factors appearing in denominators  account
for the  averaging   over the  time-like spin structures.

\vskip 0.5cm
{\bf  2.3 Classification of threshold corrections}
\vskip 0.5cm

The  generalized GSO orbifold   projection,  which    selects  the
singlet states  \wrt  the orbifold   space  symmetry group,  is
represented  by the sum over the various twisted orbifold  subsectors,
$(g,h)=(m,n)$, performed  jointly with the sum
over the  spin structures
$(\a , \b ), \quad (\a ', \b' )$
for the fermionized fields associated with the gauge degrees of freedom.

The  summations  over twisted subsectors
$(m,n), (\a , \b ), (\a ' , \b ')$  are
weighted by  phase factors  $ \e (m,n) $ and
$\eta(m,n; \a ,\b;\a',\b') $
 which are determined by the requirement that
 $\tau_2B_a $ be invariant  under
the  modular
$SL(2,Z)$ group,  generated by  $S: \tau \to -{1\over
\tau } $ and   T: $\tau \to \tau +1 $.
The set of  twisted $(g,h)$  subsectors  are mixed together under
the action of the modular group according to the
transformation   law [19,22]:
$\tau \to (a \tau +b) /(c\tau +d) ; (g,h)\to (h^c g^d , h^a g^b), [a,b,c,d
\in Z, ad-bc =1]$. (For $Z_N$ orbifolds, $ S: (m,n)\to (N-n,m), \quad
T:(m,n)\to (m, m+n)$.)
The entire set of twisted subsectors  can be organized
into disjoint  subsets   (orbits) of subsectors
which close  under the  modular
group action. The inter-orbit phase factors  $\eta (m,n,...) $
are  fixed uniquely by
the requirement of modular invariance.
The intra-orbits  (discrete torsion)
phase factors  $\e (m,n) $ are independently fixed by
constraints derived from
higher string loops modular invariance   or from  unitarity [23].
The additional freedom that might be present when the factors
 $ \e (m,n)$ are non-trivial phases   serves
then  to  label  distinct string  theories constructed from the same orbifold.
Orbifolds with no  $(g,h) $ fixed 2-d torus
(i.e., not simultaneously fixed  by
both space   $g$ and  time  $h$ twists) possess  one modular orbit only.
Orbifolds  having one simultaneous  $(g,h)$  fixed  2-d torus
possess   several modular orbits which are in
correspondence with  the distinct
$  N=2$ suborbifolds of the initial orbifold.

The multiplicity  factors $\chi (m,n) =\chi (g,h) $  count,
for twisted subsectors,   the  number of distinct   degenerate
subsectors associated with fixed points of the orbifold point group
which are  simultaneously  invariant
under   both  $g$ and $h$ [24]. (Useful information
on   these factors is provided in refs.[25,26]).
For untwisted sectors $(m=0)$, there occurs corresponding non trivial
factors $\chi (1,h)$ from the projection on
oscillator states symmetric \wrt the  orbifold point group. These
can be explicitly  calculated from the formula:
 $\chi ( 1 ,\t^n)=
\prod_i \vert -2i\sin (\pi n v_i)\vert^2= \vert det' (1-\t^n)\vert ,$
where the product and determinant  are understood to extend over the
rotated 2-d tori planes.

In the presence of Wilson lines, an additional summation
must be included over the independent Wilson lines $ a_a  $
satisfying the property $ \t ^k a_a \ne a_a$
and over the independent noncontractible loop parameters labeled
by $ m^a$. The
overall sum  over
twisted subsectors  in eq.(9) is then
replaced as, $\sum_{m,n}= \sum_{a_a} \sum_{m,n, m_{m, f}^a }$.
For  the abelian direct products orbifolds,
$Z_N \times Z_M, [M= p N, p\in Z ] $ straightforward extensions
of the above  rules  apply in which
one  deals with   pairs of
generators, $(\t_1, \t_2 ), $
shift vectors, $(v_1, v_2), (V_1, V_2), $
twisted subsectors, $(g_1  g_2; h_1 h_2)=
(m_1 m_2; n_1 n_2)$,  setting the discrete torsion phase factor as [14,23],
$\e (m_1 m_2, n_1 n_2)=e^{2\pi i k (m_1 n_2
-m_2 n_1 )/N }, \quad [k=0,\cdots, N-1]$.

The inter-orbit  phases
$\eta(m,n; \a ,\b;\a',\b') $
 depend, of course, on  the conventions
adopted
for the  fermionic determinants.
The following carefully chosen
phase conventions  for theta-functions [23,27],
  $$\eqalign { det \dh_{\a   \b }\bigg [{\t  \atop  \phi }\bigg ] =&
  \vt_{\a \b }
 \bigg [ {\t \atop
\phi } \bigg ] (\nu  =0 \vert \tau  ) = e^{-i\pi \t (\phi +2\b )}
\vt \bigg [{\a +\t \atop    \b +\phi }\bigg ](\nu =0 \vert \tau ) , \cr
\vt  \bigg [{\t' \atop   \phi' }\bigg ]
(\nu \vert  \tau )=& \sum_{n\in Z}q^{(n +\t' )^2/2
} e^{2\pi i (n+\t' )(\nu +\phi' )} ,} \q (10) $$
which describes the  determinant
of  a free complex Weyl field  obeying the boundary conditions specified
a few paragraphs above,
is found to reduce the  modular invariance constraints on the coefficients
to the remarkably simple solution  of unit phases,
$ \eta (m,n; ...) = 1$. To prove this statement
in the orbifold case,
one can  follow the same steps as in  [27] involving the use of the
identities relating  the
fermionic  and  bosonic representations  of
theta-functions and of  the  Poisson formula transforming the summation over
the compactification lattice to that  over  its dual.
Combining in this way  the  fourth  (gauge sector)
and fifth  (zero modes) factors in eq.(9)
yields an equivalent representation for the  product of these factors
in terms of
 a manifestly  modular invariant sum  over
an even,  self-dual
(shifted) (22,6)-dimensional  Lorentzian lattice,
$$ Z= \sum_{w\in \L_6 , p \in \L_6^\star  , W\in \G_{8+8} }
q^{P_L^2/2} \bq^{P_R^2/2} ;$$
$$P_{L,R}=[p_{L\mu }, P_I=W_I+k\tilde V_I; p_{R \mu } ] , \quad
\quad p^{L,R}_\mu =\pm G_{\mu \nu }w^\nu  +\ud (p_\mu -k_\mu ), $$$$
k_\mu =2B_{\mu \nu } w^\nu +P^I A_\mu ^I +\ud A_{I\nu  }w^\nu A_\mu ^I,
\quad   [p^2=p_\mu G^{\mu \nu } p_\nu ],  \q (11) $$
where $ w_\mu = \ud G_{\mu \nu  } (p_L^\nu -p_R^\nu ) =
m^ae_a^\mu , p_\mu =p_{L\mu } +p_{R\mu } = n_ae^{\star a}_\mu ,
 [m^a, n_a=$ winding  and momentum  modes integers],
 $ G^{\mu \l }G_{\l \nu }= \d^\mu _\nu $ and the basis vectors norms
 $\sum_\mu (e_a^\mu )^2 $ identify with the compactification radii $R_a$.
The  background metric and antisymmetric tensor fields,
$(G_{\mu \nu } , B_{\mu \nu }) =( G_{ab}, B_{ab})
e^{\star a}_\mu e^{\star b}_\nu , $
  and the Wilson line  vector field,
$ A_\mu ^I= a_a^I e^{\star a}_\mu $,
 represent the generalized \ccs of  the
\ws sigma model  of the heterotic string
whose action (specialized to the superconformal
gauge) is reproduced below, for definiteness,
$$ S= -{1 \over 4\pi  \a ' } \int \int  d\s d t \bigg [
\sqrt h h^{\a \b } \bigg (\dh_\a X^\mu \dh_\b X^\nu +i\bar \psi_R^\mu
\rho_\a \nabla_\b \psi_R^\nu  \bigg )G_{\mu \nu } (X)  $$$$
+\e^{\a \b } \bigg ( \dh_\a X^\mu \dh_\b X^\nu  B_{\mu \nu }(X)
+\dh_\a X_L ^\mu \dh_\b F_I A_\mu^I (X) \bigg ) -\a' \sqrt h R^{(2)} D(X) \bigg
]
 ,\q (12)$$
where $\nabla_\a \psi^\nu =\dh_\a \psi^\nu +\Omega^\nu_{\l \mu }\dh_\a X^\l
\psi^\mu , [ \O = $ generalized spin connection
\wrt to the metric and torsion
tensors]  and
$ D(X)= -\ud \ln S(X) $ denotes the dilaton field. The \sig
 background fields  in orbifolds, as in toroidal manifolds,
 are $X-$independent constants, due to the vanishing curvature tensor.

The  charge generators $Q_a$ in eq.(6)
identify with the  zero modes components  of the
Lie algebra $ G_a$ gauge current  vertex operators,
$Q_a= J_a^0 \equiv \int{d^2z \over 2\pi i} J_a(z)$.
The allowed currents are chosen among the linear combinations of
the vertex operators, $\{ i\dh F^I (z), e^{i P_I F^I (z) }\}$, invariant
under the orbifold group.
Any choice of component
$Q^{\a }_a  [\a =1, \cdots, dim  (G_a) ]$,
is admissible since
all the components  squared  $Q^{\a 2}_a $
 contribute equally
to the trace over string states.
It is easiest to work with the Cartan subalgebra generators
because of the simpler structure of their  representation as
linear combinations of the
momentum operators,  $Q_a =  Q_{aI} \int {d^2z\over 2\pi i}i \dh F^I $
with coefficients $Q_{aI}$   such that  $ Q_{ai} = \sum_IQ_{aI} E^I_i,
\quad (E^I_i , E^{\star i}_I ,  [i=1,\cdots  , 16]  $ are the
moving orthogonal frames basis and its dual for
the $\G_{8+8} $ torus)
represent
the directions  (flat components) in the $E_8\times E'_8$ weights
lattice invariant \wrt the orbifold group subject to the constraints,
$Q_{aI} V^I, Q_{aI} a_b^I \in Z$.
The weight lattice vector components representing the eigenvalues
of the Cartan subalgebra operators, $Q_a^\a , [\a =1, \cdots , rank (G) ]$,
for the momentum eigenstates, $ \vert P^I=W^I+k\tilde V^I > $, are given by
the scalar products: $ \{ Q_a^\a \cdot P =Q_{aI}^\a
P^I \}$. These relations can be used to explicitly determine the
$Q_{aI}$, their  absolute normalization  being fixed by reference to the
normalization condition, $Tr(Q_a Q_b)=\ud c(R) \d_{ab}$, for the
associated matrices.

 For non-abelian subgroup factors, the gauge group
 shift embedding case,  to which we have
 limited our considerations  here, always leads to $k_a=1$.
 For abelian subgroups, the  parameters $k_a$, which are
 still called levels for convenience of language,
 depend on the normalization of the corresponding
 charge operators $Q_a$  and specified  by [14]:
 $k_a=2\sum_I(Q_a^I)^2 $.

The insertion of the charge squared  operators is
accounted for, in the notations
introduced  in eq.(9),
by replacing the   theta-function  factors by
modified ones using the
following rule:
$$\prod_I \vt_IQ_a^{I2} \to \sum_{I\ne J=1}^8 \ Q_a^I Q_b^J
\vartheta_I' \vartheta_J'
\prod_{K\ne I,J}\vt_K+ \sum_{I=1}^8
(Q^I_a)^2\vartheta^{''}_I\prod_{K\ne I}\vt_K, \q (13)$$
where the primed and double-primed  theta-functions
are defined in terms of the sum representation given in eq.(10)
by  inserting   linear and quadratic powers of
the lattice momenta
according to the  prescriptions: $$\vt =\sum_P q^{P^2/2}, \quad
\t'^I =\sum_P  P^Iq^{P^2/2}, \quad \t''^I= 2q{d\over dq} \vt^I =
\sum_P P^{I2}q^{P^2/2}, \q (14) $$
using   self-evident shorthand  notations.
Note that the precise definition
of the 4-d chirality operator, introduced after equation (6),
reads in these notations,
 $$ \chi^2
=2 q {d \over d q } \ln { \vt \over  \eta } +{1\over 12} =
+{1\over 12} +{\vt^{''} \over \vt } +2\sum_{n=1}^\infty {nq^n\over 1-q^n}. $$
The rules  in eqs.(13) and (14)  follow  directly
from a consideration of the  bosonic representation
of  the partition  function, as  described above in connection with eq.(11).
We caution,  however,
that these rules   are not sufficient by themselves in dealing
with cases involving massless charged  oscillator states.
 For these fortunately rare  cases, one needs to
insert proper  correction factors in order to ensure a  correct
normalization of the \bet  parameters.

Turning now to
the \tres  as calculated from eq.(9) we note that the $ \D_a$ have a natural
additive decomposition  in terms of  moduli dependent
and independent contributions which we associate to
the first and second terms in
the formula: $$ \D_a (M,\bar M) = \d_a +\D_a^{(m)} (M, \bar M).$$
This separation arises when one classifies  contributions
according to  the number $N=4,2,1$  of space-time   supersymmetries
which are  realized
in terms of disjoint  subspaces of the Hilbert space
of states [3]. There exists a  one to one   correspondence  between the
\susy irreducible  representation
spaces and    the  spaces of  states of  suborbifolds which are
constructed from
subgroups of the full point symmetry group,  themselves  identified with
the modular orbits.
The $N=4,2,1$ supersymmetries are then associated with
the suborbifolds leaving fixed 3, 1 or 0  2-d tori, respectively.
The $N=4$ \susyq subsector  arises  from the purely
toroidal, trivial   orbit, $(g,h)=(1,1)$, which is clearly
absent in orbifolds, due to the projection.
The moduli dependent terms originate from N=2 suborbifolds (one
fixed 2-d torus) subsectors and  the moduli independent ones
from the  N=1  suborbifolds (no fixed 2-d torus) subsectors.
 The $N=1,2  $ orbits
generally contribute  to both
$b_a $ or $\D_a$
while  the  $N=4 $   toroidal  subsector $(g,h)=(1,1)$ (three fixed
2-d tori) contributes to neither.

The moduli dependent  $N=2$  contributions arise necessarily from
subsectors having non-vanishing
momenta, $p_{L,R}$. Indeed, a     non-trivial zero modes
factor  different from
unity occurs   only for  twisted subsectors $(m,n)$  with
a simultaneous fixed 2-d torus.
For this case, the  factors in the partition function  in eq.(9) multiplying
the zero modes factor
combine  into  the
product of an  holomorphic  function of $\tau $ times an anti-holomorphic
function  of $\bar \tau $  which, being non singular modular functions,
must   therefore
both reduce to constants  independent of
$\tau , \bar \tau $. The  modular   integral  over
the zero modes factor can then  be expressed
by  a general  formula  involving  automorphic functions for the
moduli fields  associated to   the   fixed 2-d torus.
For decomposable 6-d  tori, one finds [3]:
$$\Delta^{(m)}_a (T,\bar T)= \sum_{G'}
\sum_{i=1}^3 (\tilde b^{'i}_a)_{G'} \ln
[(T_i+\bar T_i)\vert  \eta (T_i)\vert ^4  ] ,\q (15)$$
where the sum over $G' $ runs over the distinct $ N=2 $
suborbifolds $G'$  or modular orbits and the coefficients  $\tilde b^{'i}_a $
denote the associated massless modes \bet
slope parameters   multiplied
by the ratios  of point groups orders,
$ {\vert G'\vert \over \vert G \vert }$.
The dependence on the  Dedekind function reflects
the target space duality symmetry under the $SL(2,Z)$ modular group.
The model dependent  coefficients  can also be represented as [5]:
 $\tilde b_a^{'i}\equiv  b_a^{'i}-k_a\delta_{GS}^i $, with $b^{'i}_a
=\ud [c(G_a)-\sum_{R^\a} (1+2n^i_\a )c({R_\a }) ]$,
where  $n_\a ^i $ are the
massless modes modular weights
and $\d^i_{GS}$  the
coefficients of the anomaly cancelling
Green-Schwarz counterterm. The  splitting
$ b^{'i}_a=\tilde b^{'i}_a+k_a \d_{GS}^i$
exhibits the characteristic  property  of the mechanisms responsible for the
cancellation  of
the sigma model duality symmetry  anomalies
(proportional to $b^{'i}_a$), which involve    both
\tres  ($\tilde b_a^{'i} $) and  a  gauge group  independent
Green-Schwarz counterterm  corresponding to  a  one loop
redefined  dilaton field, $S+\bar S \to S+\bar S+
\sum_i {2\d^i_{GS}\over (4\pi )^2}
\ln (T_i+\bar T_i )$.
For non-decomposable tori, the target space   modular symmetry
is  lowered  to  subgroups of
$PSL(2,Z)$.  Similar expressions  to eq.(15) continue to  hold, differing
by a non-trivial dependence on the sets of allowed moduli,   in
particular, involving  rescalings such as
$ T_i\to T_i /3 $  or $  T_i/4 $ [28].

The  moduli independent contributions $\d_a $
are associated with  the
vanishing of all components of the
momentum and winding modes, $p_{L,R}^\mu $,
yielding  therefore  a trivial zero modes factor  equal to
unity.  No analytic simplification for the modular
integral is  known to exist  in this case, for  which one must resort to
a numerical evaluation. This task is the subject of next section
and represents  the main new result reported
in this paper.

\vfill \break
{\bf 3. NUMERICAL RESULTS }
\vskip 1 cm

Before presenting the results we digress to describe how
we deal with the
 numerical  integration   over  the complex parameter $\tau  $.
The  two dimensional modular  integral can be separated in two ways:
$$\eqalign {  \int_F d^2 \tau  f(\tau_1, \tau_2)= & \int_0^\ud d\tau_1
\int_{(1-\tau_1^2)^\ud } ^\infty d\tau_2 \bigg (f(\tau_1,
\tau_2)+ f(-\tau_1 , \tau_2) \bigg ) \cr &=
\int_{\sqrt 3/2 }^\infty d\tau_2 \int_{Re(1-\tau_2^2)^\ud } ^\ud d\tau_1
 \bigg (f(\tau_1,\tau_2)+ f(-\tau_1 , \tau_2) \bigg )  .}
\q (16)$$
The general structure of the integrand is  that of an infinite sum of terms
involving products of functions
of $q , \bar q$ reading schematically,
$$B_a(\tau )= \sum_{\l , \mu }c_a (\l , \mu ) \phi_\l (q) \phi_\mu (\bar q)
=\sum_{h_L, h_R} w_a (h_L, h_R) q^{h_L} \bar q^{h_R}.$$
The projection on  the  modular group invariants  is an
essential element here  in cancelling the terms with negative powers
of ${q \choose \bar q}= e^{\pm 2\pi i \tau_1 -2\pi \tau_2}$, thus
leading to  non singular expansions with  powers identified with
the conformal weights,  $h_L= N_L+{P^2\over 2}+E_0-1,
\quad h_R= N_R +{r^2\over  2}+E_0-\ud , [E_0=\ud \sum_i [kv_i](1-[kv_i]) ,
\quad 0< [kv_i] <1 ]$.
The  functions of $ \tau_2 $ obtained upon
integration over $\tau_1$, as exhibited by the second equation in eq.(16), have
 discontinuous derivatives at $\tau_2=1$, as illustrated in figure 1.
When $h_L, h_R$  take integral values, the $\tau_1$
(Fourier) integral  for $\tau_2\ge 1$
extends over one period and so selects  the
terms $h_L=h_R$. The untwisted sector contributions have  this property and
thus reduce  for $\tau_2\ge 1$ to  constants.
The twisted sectors contributions allow  (positive) rational
values  ${k\over N }$ for  $h_L, h_R $   and  so result
in decreasing  exponentials
of the form,   $ e^{-{2\pi k\tau_2 \over N }}$.
Once the
constant parts in the full integrand,
which are  identified with the  massless modes contributions
given by $b_a$,  are
removed,    the  subtracted integrands  $(k_aB_a-b_a)$
are  fastly convergent functions.
A  cut-off
at,  say,  $\tau_2 = 2.2 $
is more than sufficient to retain the dominant part of the quadrature.
Nevertheless,  the projections  involved in the summation over the orbifolds
subsectors  cause strong
cancellations which  adversely affect the accuracy of final results.
 The  most   appropriate   way to organize calculations here  would be to
express analytically the integrand in  power expansions in
$q , \bar q$ prior to the
numerical integration [8,9]. However, this procedure is
difficult to implement in a  systematic way.
We have chosen instead  to perform  all  calculations
by brute force numerical means  and
convinced ourselves by various cross checks
 that one could maintain a  numerical
 accuracy  better than  $ 10^{-2}$ for orbifolds  $Z_N $ or
 $Z_N\times Z_M$, provided that $ N,M \le 6$, since the rounding
 errors grow with the orbifold order.
The numerical integrations are carried out in the order indicated by the
second equation in (16).

Let us  quote here useful results  concerning the inputs for some of
the orbifold parameters. Details  regarding the gauge symmetry groups and the
massless spectra can be found by consulting
refs.[14,24,25].
For the  $Z_{3,7} $ prime orbifolds,
the degeneracy factors $\chi (g,h)$ count the number of fixed points.
Thus, for twisted sectors,   $\chi (g,h) =- 27, -9,   -3 ,-1 \quad [g\ne 1 ]$,
independently of $ [h=1, \cdots , \t^N ] $,
for the $Z_3$ orbifolds with $ 0, 1,2,3 $ inequivalent
Wilson lines, respectively.
 For the $ Z_7$ orbifolds, $\chi (g,h)= -7 (-1)$,
independently of $ [h=1, \cdots , \t^N ] $,
where the first (second) numbers
refer to cases without (with)  Wilson lines.
(The reduction of the degeneracy factors in the presence of  Wilson lines,
reflecting the distinguishability of subsets of twisted subsectors, is
compensated by  a summation over the winding numbers, $m^a_{m,f}$.)
In  the $Z_4$ orbifolds, in the absence of Wilson lines,
$\chi (\t ,\t^{[0,1,2,3]})=-16,
\quad \chi (\t^2 ,\t^{[0,1,2,3]})= [16,4,16,4] $.
In  the $Z_3 \times Z_3 $ orbifold with one Wilson line
associated with the  first factor, as in the example presented below,
$ \chi (g, h) = [ 3,3,3,-9,3,3,3,-9] $
for $g=[\t_1, \t_1^2, \t_2, \t_1 \t_2 , \t_1^2\t_2, \t_2^2, \t_1\t_2^2,
\t_1^2 \t_2^2 ]$, independently of $h=\t_1^{n_1} \t_2^{n_2}$.
Note that $ \chi (1,h)=\vert  \chi (h,1)\vert $ and
$\chi (\t^m , h ) = \chi (\t^{N-m} , h)$.
The  minus signs in the degeneracy factors are inserted above
in order to  account for  a  twisted sector  dependent
phase  factor associated with the chirality.

The $N=2$ subtwisted sectors associated to  given
$(g,h) $ simultaneously  fixed planes,
consist in the $Z_4$ orbifold case
of a single modular  orbit   ${\cal O}$  of $(g,h) $ sectors
given by:  $  {\cal O}= \{ (1, \t^2),
(\t^2,1), (\t^2,\t^2) \} $, and in the $Z_3\times Z_3 $ orbifold case
of   three orbits ${\cal O}_i$,  associated
with the three fixed planes,
given by:  $$ {\cal  O}_1=\{ (1,\t_2^{1,2}), (\t_2, \t_2^{0,1,2}),
(\t_2^2,\t_2^{0,1,2})\} ; \quad {\cal  O}_2=   O_1 [ \t_2 \to \t_1 ]; $$$$
{\cal O}_3= \{ (1,\t_1 \t_2^2), (1 , \t_1^2 \t_2), ({ \t_1 \t_2^2
\choose \t_1^2 \t_2 }, 1), ({\t_1 \t_2^2 \choose \t_1^2 \t_2 }, \t_1 \t_2^2)
, ({\t_1 \t_2^2 \choose \t_1^2 \t_2 }, \t_1^2 \t_2) \}.$$

 We present our  results for   three  cases  associated with
standard embedding (2,2) orbifolds   in Table 1.
Results for  four non-standard embedding  $(0,2)$ orbifolds
are presented in Table 2. Details concerning the gauge group and the
massless spectra can be found in the second reference in [20] and in ref.[5].
Finally, to elucidate the  r\^ole of discrete Wilson lines,
\tres results for four realistic cases of
orbifolds with   three chiral matter generations
are  presented  in Table 3.
Cases A-C refer to $Z_3$ orbifolds. Up to extra $U(1)$ factors,
the   observable sector  gauge group for
Case A [14] coincides with the \sm  gauge group,
while  that for  Case B,  also
due  to Font et al. [14], is a left-right chirally symmetric gauge group
extension,
$SU(3)_c\times SU(2)_L\times SU(2)_R$ and
that of Case C, due to Kim and Kim [29], is
an intermediate unification gauge  group
$SU(3)_c\times SU(3)_w $.
Case D in Table 3  refers to a  $Z_3\times Z_3$ [14]
orbifold with an observable
sector gauge group
$SU(3)_c\times SU(2)_L\times SU(2)_R\times SU(2)$.

One of the first  calculation
of the moduli independent  \tres   was that attempted  by  Kaplunovsky  [1]
for  the simplest case of  standard embedding orbifolds.
He reported  a small   gauge group dependent term,
$\Delta =- {\D_a -\D_b \over b_a -b_b } \simeq 0.07$.
The $Z_3$ orbifold  case with two    Wilson  lines,
designated in Table 3 as  Case A,
was recently considered
by Mayr et al., [8].  Assuming
tentatively  the following decomposition
 $ \d_a=-b_a \D +k_a Y$, with the first
component   proportional  to the
factor groups   slope parameters
and the second  to  the affine levels,
 they find:
 $\D\simeq  0.079, \quad Y\simeq 4.41$.
 As for the comparison with  the existing estimates
 made in fermionic constructions of
 4-d superstrings,  this is not  very teaching because the \tres in the
 models discussed
 in ref.[6] ($\D (SU(5)) -\D  (U(1))=-24$) and
 in ref.[9] ($\D (SU(3)) -\D  (U(1))= -2.5$)  arise from  moduli
 dependent contributions  in $N=2$ sectors
 only.  In a recent systematic study, Dienes and Faraggi [34]
 report results for several new cases. They indicate, in particular,
 that the  above quoted \tres in the flipped $SU(5)$
 case [9] must be reduced by
 a factor 3.
 Let us note here that  the models obtained in the fermionic construction
 refer to specific points in the moduli space for which one
lumps together the moduli dependent and independent contributions.

The conclusions we draw from  Tables 1-3
do not strictly agree  with those of refs.[1,6].
In our results the component $-b_a \D $   proportional
to the  slope parameters  is  much     smaller than that quoted above.
 The coefficient $\D $ is never larger than a few \%  and its
sign  and magnitude change from one group factor to the other
and also from case to case. This is clearly seen on  the corrections
$\d_a$ to $U(1) $ factors where the $\D $ component is amplified by virtue
of the larger value taken there by the  slope parameters.
The  analysis of the structure of $\d_a$  does not
quantitatively support the
conjecture made in ref.[8] concerning
a universal decomposition  into   two components
proportional to $ b_a $ and $k_a$.
We do find, in agreement with ref. [8],
a large contribution proportional  to  the affine levels
 of approximate size,  $Y\simeq 1-3$.  This is not
universal, however,  but shows rather a tendency to increase  when including
Wilson lines. We remark at this point that the cases in Tables 1 and 2
featuring significantly enhanced values of $\d_a$
for certain group factors are
precisely those  cases which involve  oscillator states
charged \wrt these   group factors. Thus, charged oscillator states appear
as the main responsible for non universal effects.

We have also examined for the $Z_3\times Z_3$ orbifold, the effect of the
discrete torsion factor, $\e(m_1,m_2, n_1,n_2) = e^{2\pi i p (m_1n_2
-m_2n_1)/N}, \quad [p=0, \cdots , N]$.
The results in Table 3 refer to the case $p=0$. Although the spectrum
and hence the slope parameters $b_a$
are known [14] to   depend on the torsion,
we  find here that the \tres remain  remarkably stable
with variable $p>0$.

\vfill \break
{\bf 4. UNIFICATION AND ANOMALOUS $U(1)$  SCALE }
\vskip 1 cm
{\bf 4.1  Threshold corrections}
\vskip 0.5 cm

In this section we examine the viability
of the perturbative superstring unification within the orbifold approach.
Let us first discuss the implications of the results obtained in Section 3
for the moduli independent threshold corrections.
Assuming the simple formula, $\d_a =-b_a \D +k_a Y$, then as already noted
in connection with eq.(4),
one can absorb  the
string \tres  into
an effective unification  scale  $M'_X$   and an effective string
\cc $g'_X$.
Since   $\d_a $
 are  of positive sign, it follows that  the moduli independent
 \tres will always result in
 reduced  effective unified \cc
 and enhanced  (reduced)  unification scale, depending on whether
the  \bet  slope parameters
 $b_a $  are positive (negative),  or equivalently,   gauge (matter) dominated.
Using the numerical values  for $b_a$ and  $ \d_a$ in Tables 1-3,  we find
very small  moduli independent
corrections to  the unification scale  and/or
coupling constant, which   attain at most a few \% .

Identifying  the  string   moduli independent
\tres obtained  here,
${ \d_a\over 4\pi } \simeq  0.4 $, tentatively  with
a corresponding   field theory  threshold correction
of  typical structure  [30],
$ \d ({4\pi \over g_a^2})=\pm O(1)  \ln {M_H\over M_X }$,
yields  for the ratio of the average heavy particle mass
to unification mass, $ M_H/M_X \simeq \ud $.
Thus, one  checks  that these  contributions
are  of the same order of magnitude as the
two loop  field theory \ren  corrections [31]. We conclude therefore that
the  moduli independent \tres
should mildly affect the \he extrapolation
of the gauge coupling constants.
More quantitatively, one can estimate the corrections
to the weak angle and color  coupling constant  by means of  the formulas [5],
 $$\sin^2\t_W(m_Z)= {k_2\over k_1+k_2 } +
{\a (m_Z) \over 4\pi } {k_1 \over k_1+k_2} \bigg [
A\ln {m_Z^2\over M_X^2} + \D_A \bigg ], $$
$$\a_s^{-1}(m_Z)= {k_3\over k_1+k_2 }
\bigg [ {1 \over \a (m_Z)} +
{B\over 4\pi }\log {m_Z^2\over M_X^2} +{\D_B\over 4\pi } \bigg ],
\q (17)   $$
 where we use the notations: $ A=-(b_1k_2/k_1-b_2),
 \quad B=-(b_1+b_2-b_3(k_1+k_2)/k_3),
 \D_A  =-(\D_1k_2/k_1-\D_2), \quad \D_B=-(\D_1+\D_2-\D_3(k_1+k_2)/k_3) $.
 Evaluating the \tres for
Case A in table 3,  using $k_1 =11/3$, yields:
 $$\d (\sin^2\t_W(m_Z))
 \simeq 1.5\times 10^{-4} , \quad
\d( \a_s^{-1}(m_Z))\simeq
2. \times 10^{-2} ,\quad [ \d \a_s (m_Z)\simeq 2.7 \times 10^{-4}]  $$
 where we have set  $\a^{-1}(m_Z)=127.9\pm 0.1 $. We see that
the corrections  are rather small and lie well within  the
 present experimental uncertainties on these parameters [31],
 $\a_s(m_Z)=0.120\pm 0.010, \quad \sin^2\t_W
 (m_Z)=0.2324\pm 0.0006 $.
 The extreme smallnes  of the effect here  is due to the cancellation
 of the  predominant level dependent component  $ k_a Y$
 in $\d_a$ in the linear combinations  appearing in $\D_{A,B}$.

Turning to  the moduli dependent  corrections  $\D_a^{(m)}$,  we note that
these are generically of opposite  sign  \wrt $\d_a$ and
so have an opposite effect on the effective unification parameters.
These contributions   become  sizeable
only to the extent that  large moduli  VEVs
and  large ratios  $\tilde b'_a/b_a $
are used,
as is  clearly  demonstrated on  the following  approximate formula,
valid for large VEVs,
$$ M'_{X} \simeq  M_{X}
\bigg [ {e^{\pi (T +\bar T)\over 6 } \over T+\bar T} \bigg ]^{\tilde
b'_a \over 2b_a}. \q (18) $$
To estimate the corrections in  eqs.(17),  one can use the approximate
formulas, $\D_{A,B}\simeq  {A'
\choose B'} (\ln (2T_R) -{\pi \over 3} T_R), $  where  ${A'\choose B'}=
{A-\d A \choose B-\d B}$  such that
${A \choose B}= {28/5 \choose 20 } $ for  the \mssm
and  $\d A , \d B$ depend on
the  \mow parameters assignments. The solutions reported in refs. [5,12]
give:  $ {A \choose B} \simeq {4\sim 16\choose 24\sim  40}, $ or
equivalently, $ {A' \choose B'} \simeq {2\sim -10 \choose
0\sim 20} .$
In order for these corrections to
$\sin^2\t_W $ and $\a_s$  to reach   an order
of magnitude higher than those found above  from the moduli independent
corrections, one needs at least,
$T_R =Re(T) =O(10) $.

\vskip 0.5cm
{\bf 4.2  Standard-like superstring unification scenario}
\vskip 0.5cm

We shall now present  an extended analysis  of the string
unification picture in which
the  \ccs scale evolution proceeds through an intermediate
threshold at $ M_A $ induced by  an anomalous $U(1)$ mechanism.
A two-stage scale evolution is considered:  An  initial
short evolution  from   $M_S $ to $ M_A$,   described  by
the slope parameters $b_a^A$ set at the values   predicted
in the orbifold  models,  followed by a wide scale  evolution from
$M_A $ to $  m_Z$ described  by the  \mssm  slope
parameters. The relevant formula reads:
$${(4 \pi )^2\over g_a^2(\mu ) } =
k_a ({(4 \pi )^2\over g_{X}^2}+ \tilde Y)  +2b_a \ln  {\mu
\over M_{A}}+2b_a^A \ln {M_A \over M_X } + \Delta^{(m)}_a
(T,\bar T). \q (19) $$
We regard the five parameters
$[ g_X,  k_1 $,
 $T$ ,  $ \tilde Y \equiv  Y_0+Y , M_A ] $, which enter explicitly eq.(19),
 as adjustable parameters.
Note that $M_X $ has a  fixed linear dependence on $g_X$ which
is specified by eq.(3).
The  compactification scale can be  tentatively
identified in order of magnitude by writing: $$M_C = {2\pi \over R }
\simeq  {M_S\over 2} \bigg ({ C_{orb }\over   T }
\bigg )^\ud \simeq {2\sqrt C_{orb} M_X \over \sqrt T } , \q (20)$$
where the compactification radius $R$
and moduli VEV, $<T>=T$ are related as
$ T ={C_{orb } R^2\over \a'
(2\pi )^2 } $,
with $ C_{orb } $
a calculable  constant of order unity [26].
For, say,  the $Z_3$ orbifold, $C_{orb}=\sqrt 3/4$.
One concludes from eq.(20) that
 $M_C/M_X \simeq 1/\sqrt T.$

A rough order of magnitude estimate for the anomalous $U(1)$
Higgs mechanism scale  $M_A$
can be obtained by
imposing the condition of a vanishing  D-term scalar potential [17],
$- D_A/g_A^2= \sum_\a Q_A^\a \vert \phi_\a \vert ^2 + {g_X c_A
\over 4\a' \sqrt {k_A } }$, for a group factor $U_A(1)$ distinguished by the
index $A$.
  (The triangle anomalies  coefficient $c_A$  is defined as
  $48\pi^2 c_A= Tr(Q_A) =4Tr (Q_A^3) $, where the traces
  extend over the massless modes. This   enters  the
Green-Schwarz counterterm  through the substitution for the dilaton
field,
$ S+\bar S\to S+\bar S +c_A V_A$, whose function is  to cancel
the  various $U_A(1)$ group
factor  (gauge and gravitational) triangle anomalies, by assigning
to the gauge vector and dilaton chiral
supermultiplet fields the transformation laws, $ V_A \to V_A -\L_A -\L_A^\star
, \quad S\to S +c_A \L_A$.)
 The predicted  magnitude for the scale is:
$$ M_A \simeq  <\phi >  =
{M_P\over \sqrt {8\pi }} {g_X\over \sqrt 2}  \bigg [-{ g_X Trace(Q_A)
\over 192\pi^2  Q_{A } \sqrt {k_A}}\bigg ]^{\ud } . \q (21)$$
Using tentatively  for the model dependent ratio the estimate
$- Tr (Q_A)/(Q_{A\a } \sqrt {k_A}) \simeq 10$, one obtains:
$ M_A \simeq 1.2 g_X^{3/2} \times 10^{17 }$  GeV, which indicates  that
$M_A $ should be of the same order of magnitude as $M_X$.

We  use the
known experimental  values of the  gauge \ccs
at the  Z-boson mass, namely,  $g_1^2(m_Z)=0.127, g_2^2(m_Z)=0.425, g_3^2(m_Z)=
1.44$,  as inputs to determine  via eq.(19)
three among the above quoted adjustable  parameters.  We choose these to
be: $g_X ,\tilde Y, M_A $.
This choice is motivated by the fact that the
dependence  on these parameters
in eq.(19) can be made linear by means of   an obvious
change of variables.
 The  solutions  for $ g_X,\tilde  Y, M_A$
 are determined as a function of the
 remaining free  parameters, namely,
$ T $  and $k_1$,  and the  sets of  slope parameters, $ b_a^A, \tilde b_a'  $.
For a solution to be acceptable it must comply with the
perturbation theory
constraints  that   $g_X $ and $ Y $ be   of order
unity and with  the  obvious  inequalities
between scales, $ {M_A \over M_X } <1, \quad {M_C\over M_X } < 1 $,
which we shall eventually supplement by  the inequality,
${M_A\over M_C}<1$, reflecting the assumption that the  mechanism
inducing the scale $M_A$ is a consequence of compactification.

We  shall present the results of  numerical  applications
only for  Case A in Table 3, setting $b_a=(-11,-1,3), $ corresponding to the
\mssm , $ b_a^A
=(-71.5, -18, -9), $ as obtained from Table 3, and
$   b'_a =(18,8,6), \d_{GS}=7$, where
the  choice of slope and Green-Schwarz  parameters  $b'_a=\sum_i b_a^{'i} ,
\quad \d_{GS}=\sum_i \d_{GS}^i$ for the  moduli
dependent \tres  is based on the solutions reported in ref. [5] (see also
ref. [32]).
Regarding $k_1 $ as a free parameter when this is predicted to be $11/3$
and including moduli dependent \tres in a case  (such as the $Z_3 $ orbifold)
where these are absent,
is certainly liable to criticism. However, because the orbifold order
appears to have a minor influence on \tres and in view of the wide
freedom expected in the hypercharge gauge \cc  normalization,
we hope that  these shortcomings do not affect the consistency of our
procedure.

Our main purpose is to explain the non trivial interplay between the various
 parameters  which are most significant for string phenomenology. Choosing
 the particular subset, $k_1 , T$, as our free parameters
 while adjusting the others $(\tilde Y, M_A , g_X$) to the inputs, $g_a(m_Z^2)
 , [a=3,2,1]$ is only a technical convenience. Let us first discuss some
 qualitative features of the solutions and, in particular, the
 correlations among the parameters. The  dependence on $ \tilde Y $
 and $g_X$ shows clearly that any change in $g_X $
 can be compensated  by  a negative contribution
 to $\tilde Y $. A decrease of $k_1$ widens the distance between
the quantities  $(g_a^2 k_a)^{-1}$ and so can be compensated by
decreasing $M_A/M_X $ or $g_X$. Finally,  because the functional dependence
on $M_A $ and $g_X$ in eq.(19) involves a logarithm of these
quantities, one expects a strong sensitivity of the
parameters on  the inputs.

The results   are displayed  in figure 2.
These represent a  continous  two parameters ($k_1, T)$
family  of solutions  for $ g_X, \tilde Y, M_A$
consistent with a  \he extrapolation
of the gauge \ccs joining roughly at the common value, ${4\pi \over
g_a^2 k_a } \simeq 25$.
The physical constraints on  $\tilde Y, g_X, M_A $  select a reduced domain
for the free parameters, $k_1\in (1.4,1.8), \quad T\in (1,30)$. The
variations \wrt these parameters are monotonic. For fixed $T$, increasing
$k_1$ leads to a rapidly (algebraically) increasing $\tilde Y$ from large
negative to positive values and to less rapidly increasing
$M_A/M_X $and $g_X$. Strong variations are also found for the $T-$dependence.
However, as $T$ increases past $ T \approx 25 $,
$\tilde Y$ becomes positive and nearly
independent of $k_1$.  The values of $k_1 $ on the lower side, $k_1 <1.4$, are
excluded
by the constraints on $\tilde Y$ and those on the higher side,
$k_1>1.8$, by the constraints on $g_X $ and  $ M_A/M_X$.

A wide class of solutions occurs   with  $g_X << 1$ and
 $ -\tilde Y  >> 10^{3}$, independently of $T$ and $k_1$. These arise
 through an obvious compensation  effect  of the moduli
 independent corrections with $g_X$
 in eq.(19).
Although  the $Y_0$  component of $\tilde Y$ remains uncalculated
so far, it appears
unlikely that  this can much exceed the component $Y$ which was
evaluated in Section 3 to be of $O(1)$. In fact, since large $Y_0$
is only possible for a strongly coupled string theory involving large
$g_X $, the above must be regarded as
an inconsistent class of solutions.  (However, because the
generic dependence on coupling constant of non perturbative effects
is expected to be less suppressed in string  theory than in field theory [16],
$e^{-c/g_X}$    versus  $ e^{-(4 \pi )^2/g_X^2}$, one could possibly
achieve large $Y_0$ with not too large $g_X$.)
In the following we shall restrict  ourselves to the
conventional framework where one assumes  a smooth connection
between  string theory and its \loe limit and hence retains
the constraints $ g_X = O(1)$,  $\vert \tilde
Y \vert  = O (10) $.

Examining the variation of the solutions  with $k_1$ in figures 2(a-c),
we see that these are very rapid, especially  that of $\tilde Y$.
The condition $\tilde Y=O(1)$ can be satisfied only through a very
careful fine-tuning of $k_1$ for fixed $T$, or of $T$ for fixed $k_1$.
This is possible only in cases where $\tilde Y$
changes sign in the relevant intervals of $k_1 , T$.
The moduli dependent corrections are quite essential to achieve
  a \he extrapolation consistent with superstring
 unification.
 Incorporating the threshold $M_A$ provides solutions with reduced
 $T$.
 The constraints  on $\tilde Y $  and $g_X$
require $15 < T <30$ and $1.5< k_1 < 1.8$. Incorporating the constraint
${M_A \over M_X }  <1 $ restricts this interval to $ 1.5 < k_1 < 1.7 $.
(Narrower intervals
would be imposed if one also sets lower bounds,
say, $ {M_A\over M_X } > 10^{-1} $
and $ g_X > 10^{-1}$.) If one takes  into account the additional constraint
$ {M_A \over M_C } <1$, this would lead to the stronger bound,
${M_A\over M_X }\simeq
{M_A \over M_C \sqrt T } < {1\over \sqrt T}$, which would select  the
narrower interval, $ 1.5 <k_1 <1.6$.

For concreteness,  we
show in figure 2 (d) the scale evolution of the gauge \ccs for one  particular
solution as determined by the above procedure.
One should not be disturbed by the large value of $\vert \tilde Y\vert
$ used here, since the nearby solution  determined with a carefully tuned value
of $k_1 $ or $T$  so as to give $\tilde Y = O(1)$,
would yield nearly identical flows
for the gauge coupling constants.
This figure illustrates one of the characteristic implications of
string unification, namely, that
the  simultaneous equality at some scale of the  extrapolated \ccs  has no
special significance. The  picture depicted in figure 2 (d)
is  rather generic.  The most favorable situation  corresponds then
to an approximate joining of the \ccs  flows at a large  scale  near
$ 5 \times 10^{16} GeV $, which is
to be identified with the anomalous $U_A(1)$ scale $M_A$, associated
with the decoupling of the extra quarks or leptons modes. In the string
unification picture,
the joining scale $M_A$ can be made
larger than $ M_{GUT}$ because
of the slightly reduced normalization
of the hypercharge group coupling constant and of the spread of the \ccs
at $M_X$ which is related to the moduli dependent threshold corrections.

Let us comment briefly on  the sensitivity of the solutions to  the slope
parameters.
 (Our procedure would obviously  break down  for
$b_a^A \approx b_a$ as  this would
make the linear system of equations, eq. (19),  singular)
The slope parameters  $b^A_a$ determine
the  variation of the  \ccs
 from $ M_X $ to $ M_A$. The   choice  of $b_a^A$ is correlated to that
 of the moduli dependent
 slope parameters,  $\tilde b'_a $, since the latter  determine
the  amount by which the \ccs are spread at $M_X$.  Consider
first the case of fixed $\tilde b'_a$.
Increasing $ T$ implies a
wider spread of the \ccs at $ M_X$  which should therefore
be compensated by larger slopes $ b_a^A$ in order  to catch up with the
extrapolated
\ccs  up to $  10^{16} GeV$.
Rather than showing new plots, we only mention here that
if one  performs  a uniform reduction of the slopes
 $ b^A_a  $ by, say, a factor $2$, the solutions
 would rule out the entire domain  in $k_1, T$
 except for a narrow region around $ T=15, k_1=1.7$. Conversely, enhancing the
 slopes $ b^A_a  $ by, say, a factor $2$  ameliorates the initial picture
 without changing qualitatively the character of solutions.
 One concludes therefore that the cases involving negative
slope parameters $b_a^A$  with large absolute values (richer matter
 spectra), which are generic in orbifolds model building, are
more favorable  for unification.

The choice of $\tilde b'_a=b'_a-k_a \d_{GS} $  is also quite sensitive. Rather
than performing
an exhaustive study we have considered  two other cases  obtained from
ref. [5] and further motivated in ref. [32]. Applying  the above
procedure  of solution for these cases, we found
a significantly worsened    picture.
The first case, characterized by
  $   b'_a=(7.5,2.5,1.50), \d_{GS}= 2.5  $,  admits solutions
 only for large values of   $T > 30$ and correspondingly
 large $ k_1 > 1.8$. It improves slightly  if   reduced  values
 are used for the slopes $ b_a^A$.
 The second case, characterized by   $ b'_a=(-4.67,4, 5),  \d_{GS}= 6 $,
 admits no solutions at all,  mainly on account of an incompatibity between
 the constraints on $ Y$ and  $M_A/M_X$.  One concludes therefore
 that negative or small values for the $N=2$ slope parameters
 $\tilde b'_a $ do not constitute  a  favorable option.

Having  focused so far on standard-like compactification models, we
briefly discuss the  other two possible
classes of superstring  models.
The first refers to  compactification models  with grand unified groups,
$SU(5)$ [6]  or $SO(10)$ [21]
(up to  extra $U(1)$ factors), with a flipped assignment  for the
matter fields \wrt the standard GUT basis  or   with a
regular GUT assignment  involving higher affine levels,  $ k>1$ [33].
A perturbative weak coupling scenario  assuming a smooth evolution
from $ M_{GUT}$ to $M_X$ can be carried out in the manner described
above either  by setting
the parameters, $ b_G, \tilde b'_G $
and  $ Y $ at values specified by the  models or by imposing appropriate
constraints on them.
It should not be difficult to  obtain satisfactory solutions
 for  $ g_X $ and $ M_A$ by following a procedure similar to that used above.
 An alternative strong coupling scenario
 could also be envisaged [21] if the slope $b_G$ takes a large (gauge
 dominated) positive value and $g_X$ is large  so as to lead
 to GUT group $G$ with renormalization group invariant scale
 comparable to the string scale,
$  \L_G = M'_{X}e^{-8\pi^2k_G /b_G g'^2_{X}} $.
Although such a scenario forbids  a smooth connection from string theory
to the \loe field theory, it still provides  a
prediction  for the GUT scale, namely,
$M_{GUT}\simeq \L_G$.

The second class  of compactification models corresponds to
intermediate unification on a
semisimple  electroweak gauge group. One interesting
example is Case C in Table 3 where
the gauge symmetry at compactification,
$ SU(3)_c\times SU(3)_w\times U(1)_{P_3} $,
breaks down   to the \sm group at an anomalous $U(1)$ scale according to
$SU(3)_w\times U(1)_{P_3} \to SU(2)_w\times U(1)_Y $, where  $ Y=
T_{8w} + {P_3\over 3} $. Using the information supplied in ref.[29], we
find a level parameter $k(P_3)={1\over 3} $. This implies a
normalization of the hypercharge \cc such that $ k_1= 1+{1\over 27}=
{28\over 27} .$  Although this  falls well below the
favorable interval of $k_1$  values specified   above,
it is nevertheless interesting that the
situation for  Case C is exactly
opposite to that found above for  Case A.
A simple argument was recently made [34] that  for any (orbifold or fermionic)
model realizing a direct compactification to the standard model group,
requiring a correctly normalized hypercharge imposes the bound: $k_1 \ge {5
\over 3}$. Therefore, an intermediate unification of hypercharge in a
non-abelian group factor would appear as one of the most viable options.

\vskip 1 cm
{\bf 5. CONCLUSIONS }
\vskip 1 cm

Our results   indicate    that
the moduli independent \tres  are
comparable  in size to those for gauge field theories  in spite of the
fact that  infinitely many
massive states are integrated out for superstrings.
The  corrections  are marginally
relevant at the current  precision levels for the
\loe gauge coupling constants.
The  largest contributions  reside in
a  group independent component  $k_aY$  of size
 $Y\simeq 1\sim 3$ which remains
relatively stable with respect to  the orbifold order  or to the
choice of gauge group  embedding
and  Wilson lines.
The component $-b_a \D $ is much smaller, $ \vert \D \vert < 10^{-2}$ and
model dependent.

In order for  the  large value of the predicted  string  unification
scale $M_X$ not to  conflict
with observations, one needs
both moduli dependent \tres  (with associated compactification
scale ${M_C \over M_X} \simeq {1\over \sqrt {T } } \approx 0.3 $)
as well as  a weak hypercharge
group level parameter  varying in the narrow interval, $ k_1 =1.4\sim 1.7$.
The information that the moduli independent corrections
are $O(1)$ is useful in  providing stronger correlations among the
parameters relevant to string phenomenology.
Postulating an anomalous $U(1)$ mechanism at a
scale  $0.1 <M_A/ M_X <1$ significantly eases  the above constraints  on slope
parameters while  raising the bound on the allowed values of $ M_C$.
The resulting picture is intermediate between a delayed joining of the
\ccs flows, due
to the smaller value of $k_1$, and of  a continued flow beyond crossing,
consistent with the moduli dependent
threshold corrections.
Our analysis emphasizes the need of constructing orbifold models
combining the property of a low value for the hypercharge group
level parameter along with the usual desirable features,
namely, three chiral families, low rank gauge
group  and $ N=2 $ subsectors.
%\vskip 2 cm

\parindent=0 true cm

\def \pr  { Phys. Rev. }
\def \np { Nucl. Phys. }

\def \prl { Phys. Rev. Lett. }
\def \pl { Phys. Lett. }

\vfill \break
{\bf REFERENCES }

1.  V.S. Kaplunovsky, \np B307, 145 (1988); Erratum B382, 436 (1992)

2. P. Ginsparg, \pl B197, 139 (1987)

3. L. Dixon, V. Kaplunovsky and  J. Louis, \np B355, 649 (1991)

4. D. L\"ust, Proceedings 1991 Trieste Spring School and Workshop,
''String Theory and Quantum Gravity'',  ICTP, Trieste, Italy, eds. J. Harvey
et al.  (World Scientific, Singapore, 1992)

5. L.E. Ib\'a\~ nez,  D. L\"ust,\np B382, 305 (1992)

6. I.  Antoniadis, J. Ellis, R. Lacaze, D.V. Nanopoulos, \pl B268,
188 (1991)

7. S. Kalara, J.L. Lopez and D.V. Nanopoulos, \pl B269, 84 (1991)

8. P. Mayr, H.P. Nilles and  S. Stieberger, \pl  B317, 53 (1993)

9. L. Dolan and J. T. Liu, \np B387, 86 (1992)

10.  U.  Amaldi, W. de Boer and H. F\"urstenau, \pl B260, 447 (1991);
J. Ellis, S. Kelley and D. Nanopoulos, \pl B260, 131 (1991); P. Langacker
and M. Luo, \pr  D44, 817 (1991)

11. S. Weinberg, Proceedings of the XXVIth  International  Conference
on High Energy Physics, LLAS HEP 1992, hep-ph/9211298;
%(American  Institute of Physics, New York, 1992);
L.E. Ib\'a\~ nez, Strings 1995 Conference, Univ. of South California,
FTUAM95/15, hep-th/9505098

12. L.E. Ib\'a\~ nez,  D. L\"ust and G.G. Ross, \pl B272, 251 (1991)

13. L.E. Ib\'a\~ nez, \pl  B318, 73 (1993)

14. A. Font, L.E. Ib\'a\~ nez, F. Quevedo, A. Sierra, \np  B331, 421 (1990)

15. M. Dine and N. Seiberg, \prl 55, 366 (1985);
V. S. Kaplunovsky, \prl 55, 1036 (1985);

16. T. Banks and M. Dine, \pr D50, 7454 (1994)

17. M. Dine, N. Seiberg and E. Witten, \np B289, 589 (1987);
M. Dine and C. Lee,
\np B336, 317 (1990)

18. M. B. Green,  J. H. Schwarz and E. Witten, Superstring Theory, Vols. 1 and
2 (Cambridge University Press, Cambridge, 1987); D. Friedan, E. Martinec
and S. Shenker, \np B271, 93 (1986)

19. P. Ginsparg,  XLIXth of  Les Houches 1988  Session, "Fields, Strings and
Critical Phenomena",  eds. E. Br\'ezin and J. Zinn-Justin (North-Holland,
Amsterdam, 1990)

20. L. Dixon,
Proceedings 1987 ICTP Summer Workshop on High-Energy Physics and Cosmology,
Trieste, Italy, eds.  G. Furlan
et al., (World Scientific, Singapore, 1988)
; L. I. Ib\'a\~ nez,  XVIII International GIFT Seminar on Theoretical
Physics, "Strings and Superstrings", eds. J.R. Mittelbrunn, M. Ramon Medrano
and G. S. Rodero (World Scientific, 1985)

21. H. Sato and M. Shimojo, \pr D48, 5798 (1993)

22. N. Seiberg and E. Witten, \np B276, 272 (1986)

23. C. Vafa, \np B273, 592 (1986);  K.S. Narain,  M.H. Sarmadi
and C. Vafa, \np B288,
551 (1987)

24. L. I. Ib\'a\~ nez, J. Mas, H.P. Nilles and F. Quevedo, \np B301, 157 (1988)

25. Y. Katsuki, Y. Kawamura, T. Kobayashi, N. Ohtsubo, Y. Ono and
K. Tanioka, \np B341. 611 (1990)

26. J.A.  Casas, F.  Gomez and C. Munoz, Int. J. Mod. Phys. A8, 455 (1993)

27. P. Ginsparg, \pr D35, 648 (1987)

28. P. Mayr  and  S. Stieberger,  \np B407, 725 (1993); D. Bailin, A. Love,
W.A. Sabra and S. Thomas, \pl B 320, 21 (1994); ibid. Mod. Phys. Letters
A9, 67 (1994)

29. H.B.  Kim and  J.E. Kim, \pl B300, 343 (1993)

30. S. Weinberg, \pl B91, 51 (1980)

31. P. Langacker and
N. Polonsky, \pr D47, 4028 (1993)

32. P. Brax and M. Chemtob, \pr D51, 6550  (1995)

33. G. Aldazabal, A. Font, L.E. Ib\'a\~ nez and A. M. Uranga, University
Autonoma de Madrid preprint, FTUAM-94-28, hep-th/9410206; S. Chaudhuri,
S.-W. Chung, G. Hockney and J. Lykken, FERMILAB-PUB-94/413-T, hep-ph/9501361

34. K. R. Dienes and A. E. Faraggi, preprint IASSNS-HEP-94/113,
hep-th/9505046; ibidem, hep-th/9505018

%\vskip 1 cm
\vfill \break
{\bf TABLES CAPTIONS }
\vskip 1 cm

Table 1.  Threshold corrections for
the $Z_{3,4,7}$ orbifolds with standard gauge embeddings. The entries in the
first line are the rotation angles $\t_i,  [i=1,2,3]$ and
the shift vectors $v^i, [i=1,2,3] \quad  V^I, [I=1, \cdots , 8]$.
The second and
subsequent columns correspond
to the gauge group factors in  the observable and hidden
(primed) sectors. For each column beyond the first, the first line entry  gives
the
levels  $k_a$, the second line  gives the
\bet   slope parameters $ b_a$ or,
for the non-prime orbifolds with  $N=2 $ suborbifolds,
the pairs $(b^{N=1}_a, \tilde b'_a) $, such that $b_a=
b_a^{N=1}+\tilde b'_a $.
The third line gives the moduli independent  \tres $\d_a$.

Table 2. Threshold  corrections for
orbifolds  $Z_{3,4}$   with non standard gauge embeddings.
For each case, the  first line gives  the shift vectors  $V^I, V^{'I},
[I=1, \cdots , 8] $.
The second and
subsequent columns  correspond to a selection of the
gauge group factors in the observable and hidden (primed) sectors.
For each column beyond the first, the first line entry  gives the
levels  $k_a$, the second line  gives the
\bet   slope parameters $ b_a$ or,
for the non-prime orbifolds with  $N=2 $ suborbifolds,
the pairs $(b^{N=1}_a, \tilde b'_a) $, such that $b_a=
b_a^{N=1}+\tilde b'_a $.
The third line gives the moduli independent  \tres $\d_a$.

Table 3. Threshold corrections for  a selection of three-generations
orbifold models with two Wilson lines
(Cases A-C) and one Wilson line (Case D).
For $Z_3 $ orbifolds, the winding number parameters  attached to
the Wilson lines take the values: $ m_{k,f}=0,\pm 1$.
Case A is a \sm  group $Z_3$ orbifold model   from  Font et al.,
[14] (section 4.2):
 $3V^I=(1^42000)(20^7)', 3a_{1,2}^I=(0^72)(0110^5)',
3a_{3,4}=(11121011)(110^6)' $.
Case B is a left-right  group
$Z_3$ orbifold model  from
Font et al., [14] (section  4.3):
$3V^I=(1^42000)(20^7)', 3a_1^I=(0^72)(00110^4)',
3a_3^I=(1^321^30)(110^6)' $.
Case C  is an intermediate unification group
$Z_3$ orbifold model from Kim and Kim [29]: $3V^I=(11211200)(0^8)', 3a^I_1=
(0^311211)(1^40^4)', 3a_3^I=(0^72)(1^8)'$.
Case D  is an intermediate unification group
$Z_3\times Z_3$  orbifold
model with one  Wilson line from  Font et al., [14] (section 5):
 $ 3v_1^i = (1,0,-1), 3v_2^i (0, 1,-1);
 3V_1^I=(2110^5)(110^6)', 3V_2^I= (020^6)(0-1111000)';
 3a^{(1)I}_1=(0^511-2)(0^511-2)'. $ (The indices $1,2$ refer to the two
 $Z_N$ factors.)
For each column, the first line entry  gives the
levels  $k_a$, the second line  gives the
\bet   slope parameters $ b_a$ or, as in Case D,
the pairs ${b^{N=1}_a \choose
\tilde b'_a} $
such that $ b_a=b_a^{N=1}+\tilde b'_a $.
The third line gives the moduli independent  \tres $\d_a$.

\vskip 1 cm
{\bf FIGURES CAPTIONS }
\vskip 1 cm

Figure 1. The  threshold function $-B_a(\tau )$,
integrated over $\tau_1$,  is plotted
as a function of $\tau_2$ for the
$Z_3$ orbifold model
group factor $SU(3)_c$ of  Case B in Table 3. We  show  the
contributions of the
untwisted (continuous line)
and of the twisted  sectors (double-dashes).

Figure 2. One loop renormalization group  analysis of superstring
unification parameters based on \he extrapolation  of the gauge \ccs
starting from their  experimental
values at $m_Z$. The solutions for $ -\tilde Y$ (figure a),
$M_A/M_X$ (figure b) and $g_X$ (figure c) are plotted as a
function of $k_1$ for a discrete set of values of the moduli VEV,
$T= 1$ (continuous),
$10$ (long dash short double-dashes), $15$ (long dash short dash),
$20$ (dash dot), $30 $ (dash). The  slopes discontinuities
exibited by $\tilde Y $ in figure (a) arise because of the changes
of sign of $\tilde Y $ in this semilogarithmic plot. (For
the $T=30 $ curve, $\tilde Y >0$.) We display in figure (d)
graphs of the gauge \ccs (${4\pi \over g_a^2 k_a }, \quad [a=3,2,1]$)
variation with  renormalization scale for the
particular solution characterized by
the values, $ k_1=1.6,  T=20$, yielding the solution
$ \tilde Y=-114, M_A/M_X
=0.38, g_X=0.63$.
\vfill\eject
\magnification\magstep 1
\baselineskip =0.445cm
\def \d {\delta }
\centerline {\bf TABLE 1 }
%\vskip 0.5cm
%\settabs 5 \columns
$$\vbox {\settabs 5 \columns
%\+Orbifold \cr
\+ \cr
\+ Orbifold &$ Z_3$&  $ (113)/3 $& $(11-2)/3 $ & $(1120^5)/3$ \cr
\+ \cr
\+ Gauge Group & $ SU_3$& $E_6$  & $E'_8$\cr
\+ \cr
\+ $k_a $ &$ 1$&$ 1$& $ 1$ \cr
\+ $ b_a$  &$ -72$&$ -72$& $90 $\cr
\+ $\d_a$  &$ 2.95 $&$ 1.55 $& $1.69$ \cr
\+   \cr
\hrule
\+   \cr
\+ Orbifold &$Z_4 $& $ (112)/4$ & $(11-2)/4$ & $ (1120^5)/4$ \cr
\+   \cr
\+ Gauge Group & $ SU_2$& $E_6$& $ U(1)$ & $E'_8$ \cr
\+   \cr
\+ $k_a$ &$ 1$&$ 1$&$ 3$&$ 1$ \cr
\+$ (b_a^{N=1}, \tilde b'_a)$&  $ (-12,-42)$&$ (-36,-42)$&$ (-231,-94.5)$&$
(60,30)$  \cr
\+ $ \d_a $&$ 1.22 $&$ 1.07 $&$ 7.19 $&$ 0.77$ \cr
\+   \cr
\hrule
\+   \cr
\+ Orbifold &$ Z_7 $& $ (124)/7 $ & $(12-3)/7$& $  (12-30^5)/7$  \cr
\+   \cr
\+ Gauge Group & $ E_6$& $U(1)_1 $& $ U(1)_2 $&$ E'_8$ \cr
\+   \cr
\+ $k_a $&$ 1$&$ 4$&$ 12$&$ 1$ \cr
\+$ b_a $&$ -36$&$ -369.3$&$ -1521.$&$ 90$  \cr
\+ $ \d_a  $&$ 2.04 $&$15.6 $&$ 80.8$&$ 2.07$   \cr
\+   \cr
\hrule
}$$

%\vskip 1cm
\vfill\eject
\centerline {\bf TABLE 2 }
%\vskip 0.5cm
%\settabs 6 \columns
$$\vbox {\settabs 6 \columns
%\+Orbifold &Gauge Group \cr
\+ \cr
\+ Orbifold & $ Z_3 $ & $(1120^5)/3$ & $(1120^5)'/3$ \cr
\+ \cr
\+ Gauge Group & $ SU_3$& $E_6$&$SU'_3$& $E'_6$ \cr
\+ \cr
\+ $k_a$  &$ 1$&$ 1 $ &$1$ &$1$ \cr
\+$ b_a $  &$ -45.$&$ 9.$&$ -45. $&$ -9.$  \cr
\+ $ \d_a $  &$ 1.18 $&$ 3.66 $&$1.18  $&$ 3.66$ \cr
\+   \cr
\hrule
\+   \cr
\+  Orbifold &  $Z_3 $&  $(110^6)/3 $& $(20^7)'/3$ \cr
\+   \cr
\+  $Z_3 $& $ E_7$& $U(1)_1 $& $ U(1)_2 $& $SO'_{14}$ \cr
\+   \cr
\+  $k_a$ &$ 1$&$ 4$&$ 2$&$ 1$\cr
\+$ b_a $&$ 36$&$ -462 $&$ -105$&$  -18 $\cr
\+ $ \d_a $  &$ 3.06 $&$ 15.6 $&$$4.90 &$ 3.01 $\cr
\+   \cr
\hrule
\+   \cr
\+ Orbifold & $ Z_3 $&  $(1^420^3)/3$& $ (20^7)'/3$ \cr
\+   \cr
\+  Gauge Group & $ SU_9$   & $SO_{14}' $&$ U(1)'$\cr
\+   \cr
\+  $k_a$ &$ 1$& $ 1$&$ 2$\cr
\+$b_a  $&$ -18$&  $ 9$&$ -99 $  \cr
\+ $\d_a  $ &$ 3.64 $ &$3.66 $&$ 3.57 $ \cr
\+   \cr
\hrule
\+   \cr
\+ Orbifold $ Z_4 $&  $(1120^5)/4$& $(220^6)'/4 $ \cr
\+   \cr
\+ Gauge Group & $ SU_2$& $E_6$ &$U(1)$&  $SU_2'$  & $E_7' $\cr
\+   \cr
\+ $k_a$ &$ 1$&$ 1$& $ 12$ &$ 1$&$ 1 $\cr
\+$(b_a^{N=1}, \tilde b'_a) $&$  (-12,-42)$&$ (12,-42)$&$ (-2711,-1512)$
&$(-104,30)$&$(12,30)$\cr
\+ $ \d_a $ &$1.13$&$ 2.38$&$ 100$& $ 2.90$&$ 2.38 $\cr
\+   \cr
\hrule
}$$

\vfill\eject
%\vskip 1.cm
\centerline {\bf TABLE 3 }
%\vskip 0.5cm
%\settabs  10\columns
$$\vbox {\settabs 10 \columns
\+ $ $  Case A    \cr
\+ \cr
\+$ SU_3 $& $SU_2 $& $U(1)_1$& $U(1)_2$&  $U(1)_4$&
$U(1)_5$& $U(1)_Y$&  $U(1)'_4$&  $U(1)'_6$&  $SO_{10}'$ \cr
\+ \cr
\+$ 1$&$1$&$6$&$4$&$2$&$2$&${11\over 3}$&$2$&$4$&$1$ \cr
\+ \cr
\+$-9$&$-18$&$-227$&$-110.6$&$-31.2$&$-16.8$&$-71.5$&$-14.4$&$-69.6$&$18$ \cr
\+ \cr
\+$3.41  $&$3.41 $&$ 32. $&$ 14.4 $&$3.55 $&
$3.51 $&$ 11.4 $&$ 1.75 $&$ 13.8 $&$
3.44 $ \cr
\+   \cr
\hrule
\+ \cr
\+ $ $  Case B    \cr
\+ \cr
\+$ SU_3 $& $SU_2^L $& $SU_2^R$& $U(1)_1$&  $U(1)_2$&
$U(1)_3$& $U(1)_4$&  $SU_2'$&  $ SO_8'$&  $U_1'$ \cr
\+ \cr
\+$ 1$&$1$&$1$&$6$&$4$&$4$&$ 2$&$1$&$1$&$2$ \cr
\+ \cr
\+$-6$&$-15$&$-15$&$-216$&$-103$&$-103$&$-12. $&$-24$&$6$&$ -26.4$ \cr
\+ \cr
\+$3.57  $&$3.57 $&$ 3.57 $&$ 32 $&$14.6 $&
$14.6  $&$  -3.57 $&$ 3.56 $&$ 1.80  $&$
3.61 $ \cr
\+   \cr
\hrule
\+ \cr
\+  $ $ Case C  \cr
\+ \cr
\+$ SU_3 $&$SU_3$ &  $ U(1)_1 $& $U(1)_2$& $U(1)_3$& $U(1)_4$&
$SO_{12}' $&$U(1)'_5$&$U(1)'_6$ \cr
\+ \cr
\+ $1$&$1$ &$6$&$6$&$2$&$2$&$1$&$8$&$8 $\cr
\+ \cr
\+$ -18$&$ -18$& $ -349$&$ -284$&$ -35.8$&$ -28.6$&$ 27 $&$ -490$&$ -453$ \cr
\+ \cr
\+ $3.09$&$3.09$ & $32.8$&$ 31.4$&$ 3.65$&$ 3.49$&$ 3.13$&$ 52.8$&$ 56.3$\cr
\+   \cr
\hrule
\+ \cr
\+ $ $  Case D    \cr
\+ \cr
\+$ SU_2 $& $SU_2^L $& $SU_2^R$& $SU_3 $&  $U(1)_1$&
$U(1)_2$& $U(1)_3$&  $SU_3'$&  $ SO_6'$&  $U_1'$ \cr
\+ \cr
\+$ 1$&$1$&$1$&$1$&$2$&$4$&$ 6$&$1$&$1$&$4$ \cr
\+ \cr
\+$ {2.6 \choose -24}$&$ {-9.4 \choose -24}$&$ {2 \choose -36}$&$ {2 \choose
-12}$&
${2.9 \choose -58}$&$ {-54 \choose -112}$&$ { 2.7 \choose -270} $&${2 \choose
-12}$&
$ {2 \choose -12}$&$ { -19.6 \choose -128}$ \cr
\+ \cr
\+$1.13  $&$1.13 $&$ 1.10 $&$ 1.10$&$1.13 $&
$4.76  $&$ 9.70 $&$ 1.10 $&$ 1.10  $&$4.60$
\cr
\+   \cr
\hrule
}$$
\end